\documentclass[aps,twocolumn,showpacs,amsmath,amssymb,floatfix,superscriptaddress]{revtex4}

\usepackage[utf8]{inputenc}

\usepackage{graphicx}
\usepackage{bm}
\usepackage{color}
\usepackage{epstopdf}
\usepackage{units}
\usepackage{natbib}
\usepackage{ulem}
\begin{document}

\title{Technical advantages for weak value amplification:  When less is more}
\author{Andrew N. Jordan}
\affiliation{Department of Physics and Astronomy \& The Center for Coherence and Quantum Optics, University of Rochester, Rochester, New York 14627, USA}
\affiliation{Institute of Quantum Studies, Chapman University, 1 University Drive, Orange, CA 92866, USA}
\author{Juli\'{a}n Mart\'{i}nez-Rinc\'{o}n}
\author{John C. Howell}
\affiliation{Department of Physics and Astronomy \& The Center for Coherence and Quantum Optics, University of Rochester, Rochester, New York 14627, USA}
\date{}

\newcommand{\mean}[1]{\langle #1 \rangle}           
\newcommand{\cmean}[2]{\,_{#1}\langle #2 \rangle}   

\newcommand{\ket}[1]{|#1\rangle}                    
\newcommand{\bra}[1]{\langle #1|}                   
\newcommand{\ipr}[2]{\langle #1 | #2 \rangle}       
\newcommand{\opr}[2]{\ket{#1}\bra{#2}}              
\newcommand{\pprj}[1]{\opr{#1}{#1}}                 

\newcommand{\Tr}[1]{\mbox{Tr}\left[#1\right]}       
\newcommand{\comm}[2]{\left[#1,\,#2\right]}         
\newcommand{\acomm}[2]{\left\{#1,\,#2\right\}}      
\def\R{\mbox{Re}}                                   
\newcommand{\op}[1]{\hat{#1}}                       
\def\prj{\op{\Pi}}                                  

\newcommand{\oper}[1]{\mathcal{#1}}                 
\newcommand{\prop}[1]{\textit{#1}}                  
\def\gbar{\bar{\gamma}}
\def\ebar{\bar{\eta}}
\def\be{\begin{equation}}
\def\ee{\end{equation}}
\def\la{\langle}
\def\ra{\rangle}
\def\cur{\mathcal{I}}
\def\be{\begin{equation}}
\def\ee{\end{equation}}
\date{\today}

\begin{abstract}

The technical merits of weak value amplification techniques are analyzed. We consider models of several different types of technical noise in an optical context and show that weak value amplification techniques (which only use a small fraction of the photons) compare favorably with standard techniques (which uses all of them). Using the Fisher information metric, we demonstrate that weak value techniques can put all of the Fisher information about the detected parameter into a small portion of the events and show how this fact alone gives technical advantages.  We go on to consider a time correlated noise model, and find that a Fisher information analysis indicates that while the standard method can have much larger information about the detected parameter than the postselected technique.  However, the estimator needed to gather the information is technically difficult to implement, showing that the inefficient (but practical) signal-to-noise estimation of the parameter is usually superior.
We also describe other technical advantages unique to imaginary weak value amplification techniques, focusing on beam deflection measurements. In this case, we discuss combined noise types (such as detector transverse jitter, angular beam jitter before the interferometer and turbulence) for which the interferometric weak value technique gives higher Fisher information over conventional methods. We go on to calculate the Fisher information of the recently proposed photon recycling scheme for beam deflection measurements, and show it further boosts the Fisher information by the inverse postselection probability relative to the standard measurement case.

\end{abstract}


\maketitle

\section{Introduction}

There has arisen considerable interest in the use of ``weak value" techniques to improve the accuracy of precision measurement. While it has been recognized for some time that these techniques, in and of themselves, do not overcome fundamental limits for coherent light sources (``standard quantum limit") (see {\it e.g.} \cite{Starling2009}), there are technical advantages in that these methods make the experimental approach to these limits relatively easy with common experimental equipment. Indeed, these techniques have already been successfully applied in the lab to measure with high precision the optical spin Hall effect and other polarization dependent beam deflections \cite{Kwiat,Pfeifer2011,Zhou2012}, interferometric deflections of optical beams \cite{Dixon,Hogan,Turner}, phase shifts \cite{StarlingPhase,Feizpour2011,Xu2013}, frequency shifts \cite{Starling2010}, temperature shifts \cite{Egan}, and velocity measurements \cite{Viza}. In most of these experiments, the weak value amplification (WVA) technique met and even surpassed the sensitivity of standard techniques in the field.  For a recent review of this and related weak value research, see Ref.~\cite{review}.

Although these experimental findings have been employed in a number of different research groups and applied to metrological questions of a number of different physical parameters, there are still some open questions and even controversy \cite{Knee2013,Ferrie2013} about this technique: precisely how and to what extent can WVA techniques help against technical noise, or give some kind of technical advantage in comparison to the standard measurement techniques? Starling {\it et al.} considered a particular parameter estimator, showing that WVA could give an advantage \cite{Starling2009}. An important step was made in this question when Feizpour, Xingxing, and Steinberg \cite{Feizpour2011} were able to consider a more general kind of technical noise, and showed that so long as it had a long correlation time, WVA also help suppress it in the signal to noise ratio (SNR). In other closely related work, Kedem \cite{Kedem}, Brunner and Simon \cite{Brunner} and Nishizawa \cite{Nishizawa} also showed an increased performance of the SNR in the presence of technical noise.

In contrast to these results, recent papers have claimed that WVA gives no technical advantage \cite{Knee2013,Ferrie2013}. The argument is justified by using a Fisher information analysis of technical noise applied to the signal carrier (e.g., such as beam displacement jitter). (We note the Fisher information analysis has been recently applied to WVA by other authors as well \cite{Viza,Tanaka}.) However, this particular form of technical noise does not represent the complete picture. There are many forms of technical noise that are not incorporated in this model. For example, in optical beam deflection, noise sources include: electronics noise, transverse displacement and angular jitter, analog-to-digital discretization noise, turbulence, vibration noise of the other optical elements, spectral jitter, etc.  In light of this criticism, it is our aim in this work to analyze some of these models and examples using Fisher information and maximum likelihood methods in order to understand in precisely what sense they give or fail to give a technical advantage, as well as describe other technical advantages in beam deflection (and derivative) experiments where the imaginary WVA technique does lead to the optimal Fisher information even in the presence of some types of noise sources mentioned above.

The paper is organized as follows.  In Sec.~\ref{Fisher}, we introduce the concepts of Fisher information and maximum likelihood techniques, and illustrate how to apply them to Gaussian random measurements.  We introduce weak value amplification and postselection in Sec.~\ref{WVA}.  Uncorrelated, displacement technical noise is discussed in Sec.~\ref{KG-sec}.  Time correlated technical noise is analyzed in Sec.~\ref{FXS}.  Air turbulence is discussed briefly in Sec.~\ref{air}. The combination of displacement jitter and turbulence is discussed in Sec.~\ref{imwv}, showing the weak value technique can suppress both.  The weak value technique is shown to better suppress angular jitter in deflection measurements in Sec.~\ref{angular}.  We examine a recent photon recycling proposal in Sec.~\ref{recycle} and show the Fisher information is boosted by the inverse postselection probability.  Our conclusions are summarized in Sec.~\ref{conc}.

\section{Fisher information of an unknown parameter}
\label{Fisher}
Fisher information describes the available information about an unknown parameter in a given probability distribution. Consider an unknown parameter $d$, upon which some random variable $x$ depends. Let the probability distribution of $x$, given $d$ be $p(x | d)$. The score of the distribution is defined as $S = \partial_d \log p(x|d)$. Assuming $p$ is a smooth function, the average of $S$ over $p$ is 0, and its variance (second moment) is defined as the Fisher information,
\begin{eqnarray}
{\cal I}(d) = \la S^2 \ra &=& \int dx\, p(x | d) (\partial_d \log p(x|d))^2 \\
&=& - \int dx\, p(x | d) \partial^2_d \log p(x|d).
\label{FI}
\end{eqnarray}
Fisher information is additive over independent trials, so for $N$ statistically independent measurements (such as the collection of $N$ photons from a coherent source), ${\cal I}_N(d) = N {\cal I}(d)$.

Consider an unbiased estimator of $d$, called ${\hat d}$. This is any statistical estimator whose expectation value is $d$. The variance of ${\hat d}$ is bounded from below by the Cram\'er-Rao bound (CRB), or ${\rm Var}[{\hat d}] \ge {\cal I}(d)^{-1}$. Thus, the Fisher information sets the minimal possible estimate on the uncertainty of $d$, for any unbiased estimator.

To illustrate how this works, let us consider a Gaussian distribution for $p(\{ x_i \}|d)$, which will describe $N$ independent measurements $\{ x_i \}$ of an unknown mean with known variance $\sigma^2$ for each measurement,
\be
P_G(\{ x_i \}|d) = \prod_{i=1}^N \frac{1}{\sqrt{2 \pi \sigma^2}} \exp\left(-\frac{(x_i-d)^2}{2\sigma^2}\right).
\ee
In this example, the score is given by $S = \sum_{i=1}^N (x_i - d)/\sigma^2$, so indeed $\la S \ra =0$, and the Fisher information is given by
\be
{\cal I}_G(d) = N \sigma^{-2}.
\ee
Thus, the CRB on the variance is simply $\sigma^2/N$. Consequently, the minimum resolvable signal $d_{min}$ will be of order of $d_{min} \sim \sigma/\sqrt{N}$. In order to achieve this minimum bound on the variance, the optimal unbiased estimator ${\hat d}_{opt}$ (often called the efficient estimator) must be used. We can find it with maximum likelihood methods, by setting the score to zero, and replacing $d$ by ${\hat d}_{opt}$. In the case of $P_G$, we have
\be
S(d \rightarrow {\hat d}_{opt}) = \sum_{j=1}^N (x_j - {\hat d}_{opt})/\sigma^2 =0,
\label{max0}
\ee
so we find the efficient estimator,
\be
{\hat d}_{opt} = (1/N) \sum_{j=1}^N x_j,
\label{maxl}
\ee
which is simply the average of the data in this case. We can check that the variance gives the inverse CRB: $\la ({\hat d}_{opt} - d)^2 \ra = (1/N^2) \sum_{i,j}^N \la x_i x_j\ra = \sigma^2/N$.
In an optical context, this is the ``standard quantum limit'' scaling with $N$, which we can interpret as the photon number. Here the parameter $d$ can be interpreted as the displacement of a beam with transverse width $\sigma$. One can immediately see that the maximum Fisher information occurs for the smallest allowable beam waist. This is intuitively obvious. If a beam of very small waist experiences a small shift in its mean value, a position sensitive detector (e.g., a split detector) would see a large change in intensity as a function of its position compared to a beam with a large waist.

In the rest of this paper, we will consider coherent Gaussian distributions for simplicity. While this is somewhat restrictive, it is also quite reasonable since most of the experiments have been performed using coherent Gaussian probability densities. This is also quite nice theoretically, owing to the fact that the log likelihood function is twice differentiable and we can use variations of the simple form of the Fisher information derived above.

\section{Real weak values and postselection}
\label{WVA}
To apply the above results to recent optical experiments, we briefly recall a few facts about weak values \cite{AAV}. If a quantum system is prepared in an initial state $| i \ra$, has a system operator $A$ that is measured by weakly interacting with a meter prepared in a state of spatial variance $\sigma^2$, and then postselected in a final state $|f\ra$ with probability $\gamma = |\la f | i \ra|^2$, the meter degree of freedom will be shifted by a multiplicative factor
\be
A_w = \la f | A|i\ra/\la f |i \ra,
\label{wv}
\ee
where $A_w$ is the ``weak value'' of the operator ${ A}$, while leaving the width $\sigma$ unchanged (for the moment we consider the real weak value case for simplicity). Such behavior is in contrast to the non-postselected case, where if the initial state is an eigenstate of $A$, so that $A |i\ra = a_i |i\ra$, the average meter shift is $a_i d$, which can be much smaller in size than the weak value shift, $A_w d$.

This process gives rise to a (normalized) Gaussian meter probability distribution consisting of $N' = \gamma N$ measurement events $\{ x'_i\}$,
\be
P'_G(\{ x'_i \}|d) = \prod_{i=1}^{N'} \frac{1}{\sqrt{2 \pi \sigma^2}} \exp\left(-\frac{(x'_i- A_w d)^2}{2\sigma^2}\right).
\ee
Computing the CRB on the variance as before gives the score to be $S = A_w\,\sum_{i=1}^{N'} (x'_i - A_w d)/\sigma^2$, so the Fisher information is
\be
{\cal I}_G(d) = (A_w)^2 N' \sigma^{-2} = \la f | A|i\ra^2 N \sigma^{-2}.
\ee
Note that the post-selection probability $\gamma$ canceled out. Therefore, the Fisher information is the same as before, except for a factor $\la f | A|i\ra^2$, a number that can be arranged to approach 1 for a two-level system with a judicious choice of operator, pre- and post-selection (see Appendix). This is consistent with the SNR analysis of Ref.~\cite{Starling2009}.  Similar points were made by Hofmann {\it et al.} \cite{Hofmann}. We can then extract all the Fisher information from the weak value, showing that ideally, the WV technique can put all of the Fisher information into the post-selected events, which matches the Fisher information in the standard methods using all of the events. We note that it is not surprising that considering any sub-ensemble gives less information than the whole ensemble. What {\it is surprising} is that by using this particular small sub-ensemble gives you all the Fisher information. This fact alone gives us technical advantages, as we shall see.

Following the maximum likelihood method presented in Eqs.~(\ref{max0},\ref{maxl}), the weak value efficient estimator is given by ${\hat d}_{wv} = (1/A_w N')\sum_{j=1}^{N'} x_i$.

\section{Type 1 technical noise: Displacement noise}

\label{KG-sec}

As a next step, we include the effect of one kind of technical noise.
We first consider the technical noise model of Knee and Gauger, where the $N$ random variables $\{x_i\}$ are independently distributed with mean $d$ and known variance $\sigma^2$ \cite{Knee2013}. Gaussian white technical noise $\{\xi_i\}$ is added to $\{x_i\}$ with mean 0 and covariance $\la \xi_i \xi_j \ra = J^2 \delta_{ij}$. We note that similar models have also been previously discussed in Refs.~\cite{Feizpour2011,Starling2009,Kedem}. Knee and Gauger comment that this kind of noise might represent transverse beam displacement jitter (in a collimated beam), for example.  We call this {\it type 1} technical noise. The measured signal will be the sum of $s_i = x_i + \xi_i$, so the distribution function for $s_i$ is $p(\{s_i\}|d) = \int {\cal D}x {\cal D}\xi p(\{x_i\})p(\{\xi_i\}) \delta(s_i - x_i - \xi_i)$, where $\int {\cal D}x =\prod_{i=1}^N \int dx_i$ integrates over all variables. Integrating over $x_i$ gives the distribution of the measured results as a convolution of the two distributions,
\begin{eqnarray}
p(\{s_i\}|d) &=& \prod_{i=1}^N \int d\xi_i {\cal N} \exp\left(-\frac{(d -s_i - \xi_i)^2}{2 \sigma^2}\right) e^{-\xi_i^2/2J^2}
\nonumber \\
&=& \prod_{i=1}^N \frac{1}{\sqrt{2 \pi (\sigma^2 + J^2)}} \exp\left(-\frac{(d - s_i)^2}{2 (\sigma^2 + J^2)}\right),
\label{KG}
\end{eqnarray}
where ${\cal N}$ is a normalization constant. Calculating the Fisher information as before, for $N$ independent measurements (photons) gives
\be
{\cal I_{KG}} = \frac{N}{\sigma^2 + J^2}.
\label{IKG}
\ee
The technical noise simply broadens the width, decreasing the Fisher information.
In the weak value case, we follow the same procedure, except that $d \rightarrow A_w d$, and $N \rightarrow N' = \gamma N$, where $\gamma$ is the post-selection probability. We add the same technical noise to the resulting distribution of post-selected events $\{s_i'\}$, and get
\begin{eqnarray}
p'(\{s'_i\}|d) &=& \prod_{i=1}^{N'} \int d\xi {\cal N} \exp\left(-\frac{(d A_w -s_i' - \xi_i)^2}{2 \sigma^2}\right) e^{-\xi_i^2/2J^2} \nonumber \\
&=& \prod_{i=1}^N \frac{1}{\sqrt{2 \pi (\sigma^2 + J^2)}} \exp\left(-\frac{(d A_w - s_i')^2}{2 (\sigma^2 + J^2)}\right).
\label{KGps}
\end{eqnarray}
Calculating the Fisher information for the postselected case (which must be reduced by $\gamma$, the post-selection probability) gives
\be
{\cal {I}}_{KG}' = \frac{\gamma A_w^2 N}{\sigma^2 + J^2} = \frac{\la f |A|i\ra^2 N}{\sigma^2 + J^2},\label{IKG'}
\ee
where we used the weak value formula (\ref{wv}). Consequently, as before we find the Fisher information is modified by a factor of order 1 (which will decrease or keep the Fisher information the same, as discussed before). This is the same conclusion reached by Knee and Gauger \cite{Knee2013} and Ferrie and Combes \cite{Ferrie2013}, that the weak value amplification offers no increase of Fisher information for technical noise. This result was previously found by Feizpour, Xingxing, and Steinberg \cite{Feizpour2011}, who noted that this type 1 technical noise could not be suppressed by the real weak values technique.  However, as we will show in Sec.~\ref{imwv}, this is not the case for techniques implementing an imaginary weak value.

We go on to find the efficient estimator in the post-selected case, following the maximum likelihood method presented in Eq.~(\ref{max0},\ref{maxl}). The efficient estimator is given by ${\hat d}_{KG} = (1/A_w N')\sum_{j=1}^{N'} (x'_i + \xi_i)$. In other words, one can efficiently estimate the parameter from the SNR. We note that in either the standard or WVA scheme, the Fisher information can be further improved by reducing the width of the meter state, but only until $\sigma \sim J$, after which the technical noise dominates the variance. We will return to this point in Sec.~\ref{imwv}.

\section{Type 2 technical noise: Correlated noise model of Feizpour, Xingxing, and Steinberg}
\label{FXS}
In order to achieve the CRB, the Fisher information must not only be calculated, but the associated estimator must also be practical to implement.  The point of practicality of the efficient estimator can be strongly made by considering the analysis of Feizpour, Xingxing, and Steinberg \cite{Feizpour2011}. Consider an experiment with single photons, such that the measurement is only triggered by the photon detection at the detector, and consists of $N$ measured variables $\phi_j$ described by an average $\bar \phi$ plus a noise term $\eta_j$ that is correlated in general, $\la \eta_i \eta_j \ra = C_{ij}$ (we define this as {\it type 2} technical noise). The noise term contains both quantum and technical noise. It is then straightforward to check that the average is given by $(1/N) \sum_{i=1}^N\phi_i = \bar \phi$, and the variance is given by $V=(1/N^2) \sum_{ij} \la \phi_i \phi_j \ra$.  The two limits considered in Ref.~\cite{Feizpour2011} are (i) the white noise limit, $C_{ij}=C \delta_{ij}$, and (ii) the fully correlated limit $C_{ij}=C$. In case (i) we have $V=C/N$, while in case (ii), we have $V=C$. The SNR is given by ${\cal R}={\bar\phi}/\sqrt{V}$, thus averaging helps the SNR for white noise, but not for fully correlated noise.

Now consider the postselected case, where ${\bar \phi } \rightarrow {\bar \phi} A_w$, and $N \rightarrow N' =\gamma N$. In case (i), the variance is now $C/N'$, while the signal is $A_w {\bar \phi }$, so the SNR scales like $A_w \sqrt{\gamma}$, so the small post-selection probability drops out, indicating there is no advantage to using post-selection in this case (exactly as we calculated in the previous sections for type 1 technical noise). However, in case (ii), the signal is boosted the same amount, while the variance remains $C$ in the fully correlated case. Consequently, the SNR shows an advantage over the non-postselected case. Feizpour, Xingxing, and Steinberg go on to consider a particular noise model consisting of a combination of white and time-correlated noise, showing this advantage remains so long as the correlation time remains long compared to the photon production rate (as it is in many optical implementations).

We can now revisit this model in the context of the Fisher information metric to see how it compares to the SNR metric. The joint probability distribution of all of the variables $\phi_i$ can be written, assuming Gaussian statistics, as
\be
P_S({\bf x}) = \frac{1}{\sqrt{(2\pi)^N {\rm det} {\bf C}}} \exp\left[-\frac{({\bf x - \mu})^T\cdot {\bf C}^{-1} \cdot ({\bf x - \mu})}{2}\right].
\label{multivariable}
\ee
Here, $ {\bf x}$ is a vector of elements $\phi_i$, and $ {\bf \mu}$ is a vector of the means (in this case, $\mu={\bar \phi}{\bf 1} $ is a vector with the same mean in every element). The matrix $\bf C$ is the covariance matrix and has elements $C_{ij}$. $ {\bf C}^{-1}$ is the inverse of the covariance matrix, and ${\rm det} {\bf C}$ is its determinant.

In this case, we may calculate the Fisher information (\ref{FI}) about $\bar \phi$ contained in this distribution, and obtain,
\be
{\cal I}_S = \partial_{\bar \phi} {\bf \mu}^T\cdot {\bf C}^{-1} \cdot \partial_{\bar \phi} {\bf \mu} = \sum_{i,j=1}^N {C}^{-1}_{ij}.
\label{fi}
\ee
Notice this Fisher information contains a double sum of the elements of the inverse covariance matrix. The ``diagonal'' terms describe the self-correlation terms, and scale proportionally to $N$, recovering the independent trials for the case when the covariance matrix is diagonal. However, for strong correlations, such as the type Feizpour, Xingxing, and Steinberg consider, each element of the inverse covariance matrix can be of comparable value, so the Fisher information can scale at most like $N^2$, giving a much larger Fisher information than for uncorrelated noise. The reason is because of the correlations between the different measurements that the SNR metric misses.

When we go to the postselected case, the dimension of the matrix shrinks from $N$ to $\gamma N$, while the mean is boosted by $A_w \sim \gamma^{-1/2}$. The Fisher information is boosted by $1/\gamma$ as before, but the double sum in (\ref{fi}) now only goes to $\gamma N$ in the upper limit, ${\cal I}_S'=A_w^2\sum_{i,j}^{\gamma N}C_{ij}'^{-1}$. The covariance matrix is different in general, since it now describes the correlations between the postselected photons only. Consequently, for white noise the Fisher information is the same, up to a factor of order 1, while for highly correlated noise, the Fisher information scales at most as $N^2 \gamma$, so it is actually decreased by a factor of $\gamma$ by the weak value technique, compared to the non-postselected case. It is easy to see why: the correlations of any single postselected photon with any rejected photon are lost in the detection scheme (unless further processing of the correlated missing photons is done), and consequently cannot be harnessed to further suppress the noise. In contrast to this, if the photon is correlated only with itself, then taking a random postselection will not hurt, and the Fisher information can stay the same.

However, this is not the whole story. We must ask what estimator should be used that saturates the CRB, and whether it is practical to implement it. We can find this estimator using maximum likelihood methods described earlier (\ref{max0},\ref{maxl}) to find the estimator $\hat \phi$.  We assume that the covariance matrix is positive definite, symmetric, and invertible. It then follows that its inverse is also symmetric. We find the estimator in the non post-selected case to be
\be
{\hat \phi }_S= \frac{ \sum_{i,j=1}^N {C}^{-1}_{ij} \phi_j}{\sum_{i',j'=1}^N {C}^{-1}_{i'j'} }.
\label{estimator}
\ee
We can check the variance of this estimator matches the CRB,
\be
\la ({\hat \phi }_S - {\bar \phi})^2 \ra = \frac{\sum_{i,j,k,l} C^{-1}_{ij} C^{-1}_{kl} \la (\phi_j - {\bar \phi})(\phi_l - {\bar \phi}) \ra}{\left(\sum_{i',j'} {C}^{-1}_{i'j'} \right)^2},
\ee
because the correlation of the two random variables gives the elements of the covariance matrix, $C_{jl}$, which cancels one of the inverse matrices in the sums, giving one factor of the denominator, resulting in the CRB we found above, the inverse of (\ref{fi}).

However there is a difficulty in this result: The experiment implementing the estimator (\ref{estimator}) must multiply every data point $\phi_j$ by a different weighting factor, $f_j= \sum_{i=1}^N {C}^{-1}_{ij}/ \sum_{i',j'=1}^N {C}^{-1}_{i'j'}$, which knows about the rest of the data points. The experimenter must know exactly what the correlations are, how many data points are being collected, and must be able to weight each data point by a different factor in order to extract the maximal information in the data average. This is generally a very challenging experimental task. Therefore, the SNR, which treats every type of noise on equal footing (so the weighting assignment is $f_j = 1/N$ for all $j$) is usually the most practical option. Consequently, the SNR, although suboptimal as a means for estimating in this case, is still superior since the optimal estimator is impractical to implement for typical experiments. This is why the SNR is used by experimentalists: because a complete categorization of the noise correlations is a formidable task requiring detailed knowledge of noise correlations and extensive post-processing.  If we go on to consider non-Gaussian correlated noise, such as $1/f$ noise, the problem of implementing the estimator becomes even worse.


\section{Type 3 technical noise - air turbulence}
\label{air}
Another type of technical noise that is very important in open air experiments is turbulence, which we refer to here as type 3 technical noise. While we will not give a quantitative analysis of this effect here, it gives additional beam width broadening beyond the diffraction limit because of beam breathing (on a short time scale), and beam wander (on a longer time scale) because of the propagation through the random medium \cite{book}. This becomes an important problem when there is a large optical path length from the position where the deflection occurs to the detector where it is measured. Beam jitter from turbulence cannot be underestimated when dealing with extremely small deflections. The beam wander effects become important on time scales longer than the ratio of the beam width to the typical air velocity. Typical experiments can run between seconds and hours, so this effect must be accounted for. The weak values schemes have a distinct advantage over the standard beam deflection measurement in having short optical path lengths.  We will see this in detail in the next section.

\section{Imaginary weak values and technical advantages}

\label{imwv}

We proceed to consider situations where imaginary WVA has technical advantages for combined technical noise types. Kedem points out that in the case of imaginary weak values, noise in the average position does not appear in making measurements in the momentum basis, and that noise on the average momentum helps the SNR \cite{Kedem}. We focus on a different effect unique to imaginary weak values. For definiteness, consider beam deflection measurements for a coherent Gaussian beam using standard deflection techniques versus a Sagnac interferometer weak value experiment. Most experiments used to date that show technical advantages use imaginary weak values.

To start, assume that the system has no technical noise. The unknown parameter of interest is the deflection $k$ of the beam, which can be interpreted as the transverse momentum kick given by a mirror. Treps {\it et al.} showed that one can interfere a local oscillator of a first order TEM mode to achieve the optimal Fisher information \cite{Treps}. However, for simplicity and to understand the role of beam diameters we consider a more standard approach. A tilt or deflection on a beam requires propagation to observe a displacement. The standard method of measuring an unknown mirror tilt $k$ with a beam of width $\sigma$ is to propagate the light and focus it with a lens, thereby taking the Fourier transform of the beam by measuring the beam with a split detector in the back focal plane of the lens (see Fig. 1(a) assuming $q=0$). The lens transforms the tilt $k$ into a displacement $fk/k_0$ on the detector, where $f$ is the focal length and $k_0$ is the wavenumber. The lens takes the beam width $\sigma$ at the mirror to a beam width at the focus given by $\sigma_f = f/2k_0 \sigma$.

\begin{figure}[tb]
\centering
\includegraphics[width=8.5cm]{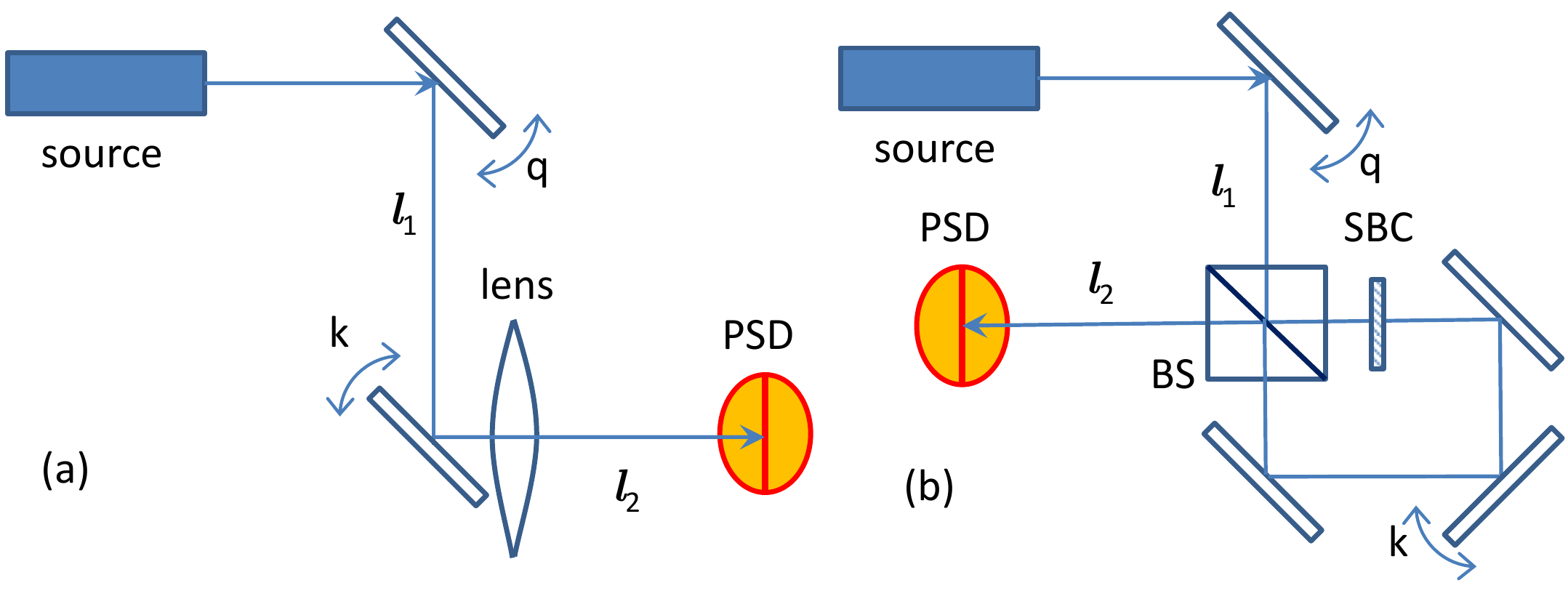}
\caption{Two strategies to detect an unknown constant momentum kick $k$ given angular jitter of momentum $q$. (a) Standard method: After the momentum kick, the beam propagates a distance $l_1$ before getting a momentum kick $k$ from the mirror, that is the parameter we wish to estimate. The beam then passes through a lens with a focus $f = l_2$, chosen to equal the distance traveled until it comes to the split detector, where the beam reaches its focus at a position sensitive detector (PSD). (b) Weak value type method: After the momentum jitter kick $q$, the light enters a Sagnac interferometer, comprised of a 50/50 beam splitter (BS), Soleil-Babinet compensator (SBC), which introduce a relative phase shift $\phi$ between the paths, other mirrors, including the one that has the momentum kick $k$ we wish to measure. The beam ends at a position sensitive detector (PSD).}
\label{ok}
\end{figure}

The single photon probability distribution function is then
\be
p_H(x| k) = {\cal N} \exp\left[-\frac{(x - f k/k_0)^2}{2\sigma_f^2}\right],
\ee
where ${\cal N}$ is a normalization. If we send $N$ independent photons to measure $k$, this yields a Fisher information of
\be
{\cal I}_H = \frac{N (f/k_0)^2}{\sigma_f^2} = 4 N \sigma^2,
\ee
which is an intuitive result. This means that in order to get the smallest beam waist in the back focal plane, one wants the largest beam waist possible before the lens. We now compare this with the Sagnac interferometer weak values result.

The weak value technique for measuring the beam deflection (Fig. 1(b) with $q=0$) gives the post-selected distribution, that when properly normalized, is a Gaussian distribution with mean $4 k \sigma^2/\phi$ and width $\sigma^2$. Here, $\phi$ is the phase difference of clockwise and counter-clockwise photons in the interferometer applied by the Soleil-Babinet compensator (SBC). This applies only to the post-selected fraction $N' = \gamma N$ of the photons that satisfy the postselection criterion, where $\gamma = \phi^2/4$. The postselected single photon probability distribution function is then
\be
p_H'(x| k) = {\cal N}' \exp\left[-\frac{(x - 4 k \sigma^2/\phi)^2}{2\sigma^2}\right].
\ee
The Fisher information for the postselected measurements, reduced by the postselection probability $\gamma$ ($N \rightarrow \gamma N$) is then found to be
\be
{\cal I}_H' =(\gamma N) 4 \sigma^4/(\gamma \sigma^2)= 4 \sigma^2 N.
\label{Fisherimaginary}
\ee
Thus, the weak values and standard method result in {\it the same} amount of classical Fisher information, as also discussed in the appendix \cite{note}.  It is important to realize that this calculation is only valid in the regime of small angles $\sin^2(\phi/2)\approx \phi^2/4$.  In other words, in the small angle or weak value regime, all of the Fisher information is in the photons leaving the dark port of the interferometer.

So how does noise affect the Fisher information for the two methods? As a first step, we will assume that there is no transverse beam deflection jitter, but only transverse detector jitter. In other words, there is a small transverse deflection $k$ with no technical noise on the beam, but the detector used to measure the beam has a transverse jitter $\xi$ that we only sample at the photon arrival times, giving type 1 technical noise. We first consider the standard method. As before, the new likelihood function is the convolution of the technical noise-free likelihood function with a Gaussian of width $J$. This simply increases the average beam waist at the detector by $\sqrt{\sigma_f^2 + J^2}$, resulting in a Fisher information
\be
{\cal I}_N = \frac{N (f/k_0)^2}{\sigma_f^2+ J^2} = \frac{4 N \sigma^2}{1 + \left(\frac{2 k_0 \sigma J}{f}\right)^2}.
\ee
This is a rather interesting result. It means that this approach can achieve the maximum Fisher information (as if there were no noise at all), but only in the limit of large focal length ($f \gg 2 k_0 \sigma J$). A large focal length is equivalent to opting for a larger displacement ($f k /k_0$) while allowing for a large focal spot ($\sigma_f \gg J$). This is once again an intuitive result: If a very small focal spot lands on a detector that has Gaussian random shifts, large differential intensity fluctuations will occur, due to detector jitter, compared with a large focal spot.

For the imaginary weak value approach, the results are quite different. The beam waist for this case is given by $\sigma$, not the focused beam waist $\sigma_f$. This gives a Fisher information of
\be
{\cal I}_N' = \frac{4 N \sigma^4}{\sigma^2 + J^2}.
\ee
Therefore, since $\sigma \gg \sigma_f$ and there is freedom to choose $\sigma$ as large as one wishes, we can make the beam waist much larger than $J$, which both suppresses $J$ and increases the Fisher information (\ref{Fisherimaginary}). This shows that Fisher information for the imaginary weak values approach remains unchanged in the presence of transverse detector jitter, while the focused beam approach requires long focal lengths.

Hence, the measurement geometry plays a very important role in several ways. Ideally one would like a detector with a continuous detection distribution, but practical considerations mean there will always be dead space between finite width detectors. A simple method for obtaining nearly optimal beam detection is split detection. However, even in this case there is always a gap between the detectors, which sets a minimum beam diameter and thus a finite propagation through horizontal turbulence. This is not a problem for weak values since the detector can be placed immediately after the last beam splitter which mitigates the type 3 technical noise while also suppressing type 1 technical noise. The standard method operating in the regime where type 1 noise is completely ameliorated will suffer from the type 3 technical noise.

\section{Type 4 technical noise - angular beam jitter}
\label{angular}
We now consider angular beam jitter by modeling it as a random momentum kick $q$ given to the beam before it enters the interferometer, or before it approaches the signal mirror in the standard method. This kick could be from air turbulence or mirror jitter.

\subsection{Standard method results for angular jitter}

We first consider the standard method of measuring the beam displacement illustrated in Fig. 1(a). The Fourier optics of this geometry is described by a series of unitary operators that act on the transverse degree of freedom in the paraxial approximation. Starting from an initial transverse state $|\psi\ra$, that we take to be a Gaussian in transverse position with zero mean, and variance $\sigma^2$, the unitary $U_q = \exp( i q {\hat x})$ gives the first random momentum kick $q$ to the beam. This is followed by a propagation of distance $l_1$, given by the unitary $U_{l_1} = \exp (- i {\hat p}^2 l_1/2 k_0)$, where $k_0$ is the wavenumber of the light. This is followed by a momentum kick $k$ described by $U_{q \rightarrow k}$, and then the lens which gives a quadratic phase front, $U_f = \exp( - i k_0 {\hat x}^2/2f)$. The final propagation $U_{l_2}$ puts the beam at the measurement device, which is a position-sensitive detector (PSD).

Taken together, we can describe the final state as
\be
\psi_d(x) = \la x | U_{l_2} U_f U_k U_{l_1} U_q | \psi \ra.
\ee
To obtain an explicit form for the state, and the expectation of the position and its variance in this state, a series of complete sets of states are inserted between the unitaries. As an intermediate step, let us define state $\psi_i$ as the state after the $k$ momentum kick, but before the lens. The action of the lens is diagonal in the position basis, but the subsequent propagation is diagonal in the momentum basis. Defining the coordinate at the detector as $x$, the state at the detector $\psi_d(x)$ is given by
\be
\psi_d(x) = \int \frac{dp}{2\pi} \int dy \psi_i(y) e^{-i k_0 y^2/2f - i py} e^{-i p^2 l_2/2k_0 + i px}.
\ee
We can reverse the order of integration and perform the $p$ integration first. This gives
\be
\psi_d(x) = \sqrt{\frac{k_0}{2\pi i l_2}} \int dy \psi_i(y) e^{-i k_0 y^2/2f + i (y-x)^2k_0/2l_2}.
\ee
Let us choose $l_2 = f$, so the quadratic terms in the exponential cancel out, which will lead to focusing the beam.  This leaves the state as
\be
\psi_d(x) = \sqrt{\frac{k_0}{2\pi i l_2}} e^{ i k_0 x^2/2f}
\int dy \psi_i(y) e^{-i y k_0 x/f},
\ee
which simply gives the scaled Fourier transform of $\psi_i$, but with a particular value of the momentum, ${\tilde \psi}_i(p \rightarrow k_0 x /f)$.

The remainder of the solution involves finding the intermediate state. This is straightforward in the momentum basis, because the first three operators are diagonal in this basis. Therefore, the momentum-space expression for ${\psi}_i$ is given by
\be
{\tilde \psi}_i(p) = (8 \pi \sigma^2)^{1/4} \exp( - \sigma^2 (p + q + k)^2 - i l_1 (p+k)^2/2k_0).
\ee
Putting these results together, we find that the final state at the detector gives a Gaussian distribution in position $x$ with average and variance of
\be
\la x \ra = - (k+q) f/k_0, \qquad \la (x - \la x\ra)^2\ra= f^2/(2 \sigma k_0)^2.
\label{y}
\ee
The net result is that the random momentum kick $q$ simply adds to the signal $k$. Averaging this distribution of Gaussian random jitter of momentum $q$ with zero mean and variance $Q^2$ gives another Gaussian distribution for $x$, of mean $- k f/k_0$ and wider variance $ f^2 [ 1/(2 \sigma k_0)^2 + Q^2/k_0^2]$. Thus, the Fisher information, for $N$ independent measurements, about $k$ in this distribution is given by
\be
{\cal I}_d = \frac{4N \sigma^2}{1 + (2 \sigma Q)^2}.
\label{idirect}
\ee

\subsection{Weak value treatment of angular jitter}

A similar analysis can be carried out for the weak value interferometer case. In addition to the momentum kick and the propagation steps, there are two unitaries that depend on the which-way operator ${\hat W} = |\circlearrowright \ra \la \circlearrowright | - |\circlearrowleft \ra \la \circlearrowleft |$, where the states $|\circlearrowright \ra, |\circlearrowleft \ra $ are clockwise and counterclockwise moving photon states inside the Sagnac interferometer. Superpositions of these states are created by the 50-50 beam splitter operating on the incoming beam. The unitaries are the relative phase shift operator, $U_\phi = \exp( i \phi {\hat W}/2)$, induced by the SBC, and the which-path momentum kick operator, $U_{Wk} = \exp( i k {\hat W} {\hat x})$ delivered by the signal mirror.

As shown in Fig. 1(b), starting in the state $|\Psi_i\ra = |\psi\ra (|\circlearrowright \ra + i | \circlearrowleft \ra)/\sqrt{2}$, and ending in state $|\Psi_f\ra = |x \ra (|\circlearrowright \ra - i | \circlearrowleft \ra)/\sqrt{2}$, we seek the transition amplitude
\be
\la \Psi_f | U_{l_2} U_{kW} U_\phi U_{l_1} U_q | \Psi_i\ra,
\ee
which describes photons entering in one interferometer port and exiting the other interferometer port.
Fortunately, the which-path states pass through all but two of the operators, so this amplitude may be simplified to eliminate the which-path states and operators, giving the state of the transverse beam at the detector, $\psi_{wv}(x)$, to be
\be
\psi_{wv}(x)= \la x| e^{-i {\hat p}^2l_2/2k_0} \sin(k {\hat x} + \phi/2) e^{-i {\hat p}^2l_1/2k_0} e^{i q {\hat x}} | \psi \ra.
\ee
The detailed calculation of this state involves inserting complete sets of states, resulting in a complicated expression. The result is simplified by expanding to linear order in $\phi$ and $k$ since we assume the weak value ordering of parameters, $k \sigma < \phi < 1$. We must renormalize the post-selected distribution by the probability $\gamma$ of a photon exiting the dark port, given by
\be
\gamma = (\phi/2)^2 + k^2 \sigma^2 + {\cal O}(k^2 q^2) + {\cal O}(\phi k q) +\ldots.
\ee
We drop the other terms compared with $(\phi/2)^2$, so the phase shift $\phi$ controls the post-selection.  This gives an average displacement of
\be
\la x\ra = \frac{4 k \sigma^2}{\phi} + \frac{k l_1(l_1+l_2)}{ k_0^2 \phi \sigma^2} + \frac{q (l_1+l_2)}{k_0} + \ldots
\ee
where we suppress higher order terms in $k, q, \phi$. The first term of order $k$ is the usual weak value term, amplified by $1/\phi$ \cite{Dixon}. The second term comes from the diffraction effects which gives a small correction to the first term because we take $\sigma^2 \gg \sqrt{l_1(l_1+l_2)}/2k_0$. The remaining term, $q(l_1+l_2)/k_0$ is just the free propagation from the momentum kick, and has a geometric optics interpretation of the imparted deflection angle $q/k_0$ times the total length of propagation. We note that the way $q$ enters into the average displacement has a very different form than Eq.~(\ref{y}). We will return to this shortly. The displacement variance at the detector may be similarly calculated to find,
\be
\la (x - \la x\ra)^2 \ra = \sigma^2 + \left(\frac{l_1+l_2}{2 k_0 \sigma}\right)^2,
\ee
plus further corrections of order $k^2$. Thus, we see that the diffraction effects broaden the width of the beam, proportional to the total path length. If we approximate the distribution as a Gaussian with the mean and variance discussed above, the Fisher information about $k$ can be found by averaging over $q$ as before to find
\be
{\cal I}_{wv} = \frac{4N \sigma^2}{1 + \left( \frac{l_1+l_2}{2 k_0 \sigma^2}\right)^2 \left( 1 + (2 \sigma Q)^2\right)}.
\ee
Consequently, even if $ \sigma Q$ is large compared to 1, so as to degrade the Fisher information in the standard method (\ref{idirect}), in the weak values technique, there is an additional suppression factor of the amount of diffraction, $(l_1+l_2)/(2 k_0 \sigma^2) \ll 1$, indicating that the weak value technique outperforms the standard method when dealing with angular jitter. This result can be understood intuitively because the angular jitter directly adds to the detected deflection in the standard measurement case, whereas it is only a small correction to the deflection that is controlled by diffraction in the weak value technique.

The rather remarkable properties associated with imaginary weak value experiments may not be entirely attributed to weak values, but to geometric terms associated with the experiments.  These terms, such as beam waist diameter, arise even when WVA is small.  However, it is important to note only in the limit of large WVA that all of the information in the measurement can be placed in the measured photons and that optimal SNR estimation of the parameter can be made.  Therefore, these geometric terms and WVA work hand-in-hand to achieve the technical noise suppression.  

\section{Fisher Information for recycled photons}
\label{recycle}
We can also consider the possible benefit to the Fisher information of the recent photon recycling proposal of Dressel {\it et al.} \cite{Dressel2013}. Considering photons as a resource, one can ask what the maximum amount of information can be extracted from a given set of photons.  In much the same way as a high finesse cavity can increase phase sensitivity in an interferometer, recycling photons from the bright port of a weak value experiment can increase the information about a parameter.  The central idea is to postselect all of the photons while keeping the large weak value amplification.  The scheme is to recycle the rejected photons by closing off the interferometer, so the remaining $N_1 = N - N'$ photons are re-injected ($N' = \gamma N)$ and once again sample the unknown parameter $k$. The authors show that it is critical that the rejected light be reshaped to once again be in its original profile, otherwise all amplification of the split detector SNR is erased over many cycles. Below, we consider the simplest version where no propagation effects are included. In that case, the second round of postselected light has exactly the same distribution on the detector, with the same post-selection probability $\gamma$, so the Fisher information for this cycle is ${\cal I}_1 =(\gamma N_1) 4 \sigma^4/(\gamma \sigma^2) = 4 \sigma^2 N_1$. This process is now repeated many times, where $N_i$ is the number of rejected photons on the $i^{th}$ round, and we define $N_0 = N$. The uniform postselection probability indicates that $N_i = N_{i-1} (1-\gamma)$, implying that $N_i = N (1-\gamma)^i$.

Since each measurement is independent from the last, the Fisher information simply adds, giving a total of
\be
{\cal I}_{\rm tot} = 4 \sigma^2 \sum_{i=0}^\infty N_i = 4 \sigma^2 N \sum_{i=0}^\infty (1-\gamma)^i = \frac{4 \sigma^2 N}{\gamma}.
\ee
Therefore, the Fisher information has been boosted by a factor of $1/\gamma$ compared to the standard method, or the single pass weak value method. Here, we are considering the photon number $N$ as the resource. Of course, as is pointed out in Ref.~\cite{Dressel2013}, one could have been sending more light onto the detector using the standard method in the time taken for the light recycling or employed other standard schemes, but there could be many technical reasons why this may be impossible or inadvisable given a laboratory set-up. There may be a minimum quiet time between laser pulses, for example, or the detector may have a low light power threshold. We note that the profile reshaping process actually removes information about the parameter $k$, but does so in a way that the estimator can be simply and easily implemented as the split detection SNR. Together with the technical advantages already discussed, this is an important improvement over previous techniques. When further combined with quantum light techniques, this method gives a powerful advantage for estimating a parameter. The fact that all photons are now collected permits one to further mine interphoton correlations for noise suppression.

\section{Conclusions}
\label{conc}
We have shown how weak value based measurement techniques can give certain technical advantages to precision metrology. First and foremost, the Fisher information in a weak value measurement (which uses a small fraction of the available light) can be as large as the Fisher information of a standard measurement (which uses all of the available light). This is remarkable because the remaining light can be sent to another experiment \cite{Starling2010}, or recycled \cite{Dressel2013} to give even higher amounts of Fisher information.

We have also explored technical advantages the weak value experiments can have over standard measurement techniques. Obvious advantages, such as when the detectors saturate at a certain light intensity, have been pointed out in previous works \cite{Starling2009,Dixon}.  The possible advantages for different types of technical noise should be investigated on a case by case basis. There are cases where the weak value techniques gives advantages, and other cases where it is at a disadvantage, and yet other cases where there is no difference.
It is clear, for example, that dephasing noise will reduce the size of the weak value, which will be detrimental to this technique \cite{Briggs}.
We have shown how detector noise, together with air turbulence can both be eliminated by a weak value deflection measurement, whereas the conventional standard method must suffer from turbulence jitter if detector noise is suppressed in open air experiments.  We also demonstrated that for angular jitter, the weak value technique for beam deflection measurements can have much higher Fisher information than the standard technique.  Considering the wide range of experiments that have now successfully employed weak value techniques to make high precision measurements \cite{Kwiat,Pfeifer2011,Zhou2012,Dixon,Hogan,Turner,Starling2009,Feizpour2011,Xu2013,Starling2010,Egan,Viza}, this should not be too surprising.

Another conclusion we reach is that it is not sufficient to show an estimator does not reach the CRB to decide it should be rejected. Rather, the optimal estimator should be found, and must be practically implementable.  If it is not, the inefficient - but practical - estimator is superior.  We have argued that time-correlated technical noise is one example where the difficulty of implementing the optimal estimator is outweighed by the option of implementing the postselection with amplification.

\begin{acknowledgments}

We acknowledge support from the US Army Research Office Grants No. 62270PHII: STIR, No. W911NF-09-0-01417, No. W911N-12-1-0263, as well as the National Science Foundation Grant No. DMR-0844899.

We thank Y. Aharonov, J. Tollaksen, G. C. Knee, E. M. Gauger, J. Dressel, and H. Wiseman for helpful comments on the manuscript.

\end{acknowledgments}

\appendix*

\section{Weak value for a two-level system}\label{qubit}
Our calculations are based on an interaction of the type $U=\exp(-i\,d\,\hat{p}\,\hat{A})$, where $d$ is the unknown small parameter. Tracking/measurements of the meter degree of freedom $x$ after the postselection gives a SNR of ${\cal R} = (\sqrt{N}\,d/\sigma)\times|\langle f|i\rangle\,\mbox{Re}(A_w)|$. If instead, measurements of the conjugate momentum $p$ are performed, the result is ${\cal R} = (\sqrt{N}\,d/\sigma)\times |\langle f|i\rangle\, {\rm Im}(A_w)|$ \cite{Kedem}. 
We will show that under the right choices for pre- and post-selection, $|\langle f|i\rangle\, {\rm Re}(A_w)|=1$ or $|\langle f|i\rangle\, {\rm Im}(A_w)|=1$.

\begin{center}
\begin{figure}[h]
\includegraphics[scale=0.4]{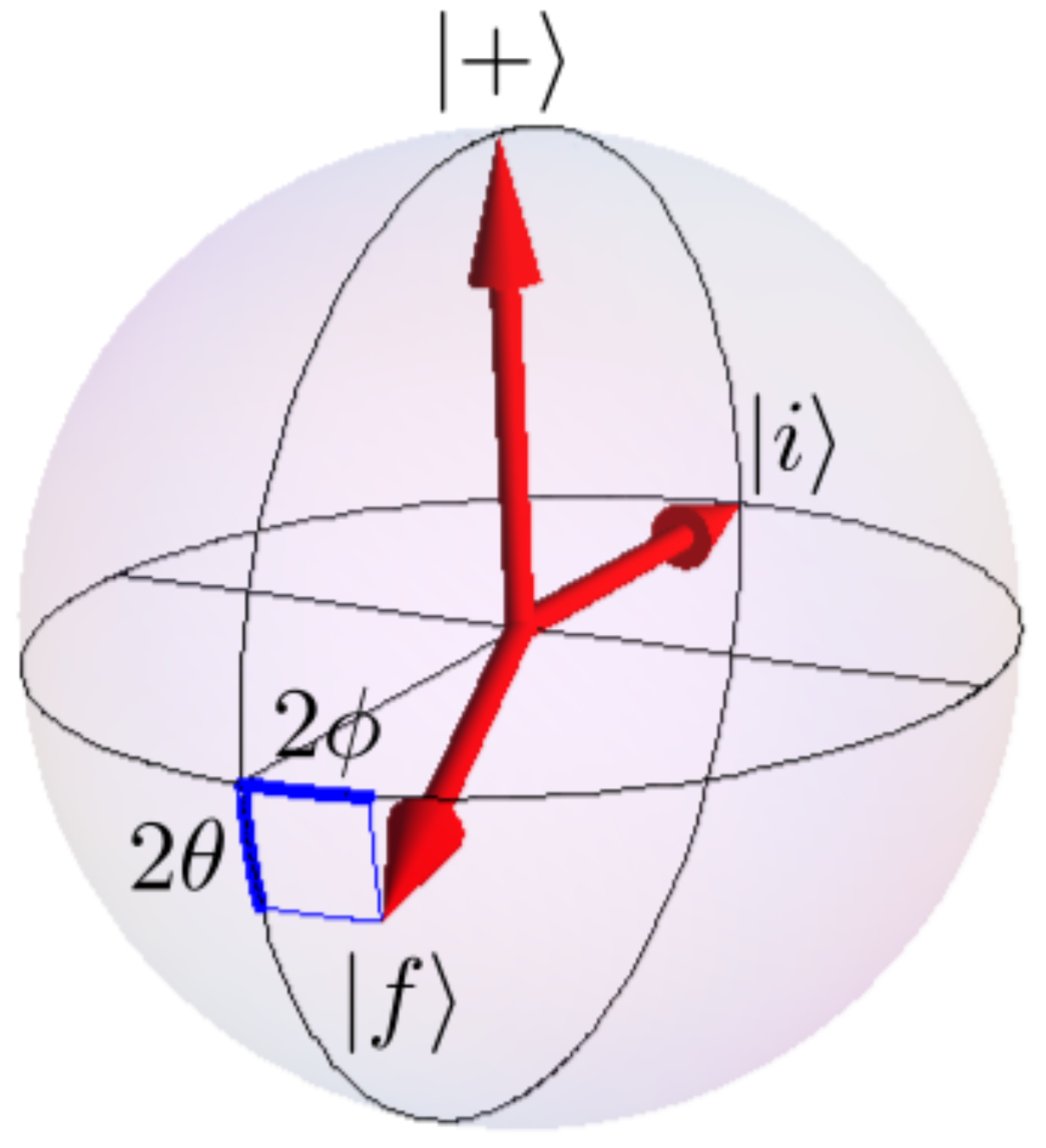}
\caption{\label{Sphere}Representation on the Bloch sphere of the preselection and postselection states for $\Theta=\pi/2$. The state $|i\rangle$ corresponds to a unit vector anywhere on the equatorial plane, and the state $|f\rangle$ is off to be antiparallel to $|i\rangle$ in the two angular directions. The angle $2\theta$ ($2\phi$) gives origin to the real (imaginary) part of the weak value.}
\end{figure}
\end{center}

We chose to consider the operator $\hat{A}$ such that $\hat{A}|\pm\rangle=\pm|\pm\rangle$, where $|+\rangle$ and $|-\rangle$ form an orthonormal basis for the Hilbert space of a two-level system. The initial state of the system can be written as
\begin{equation}
|i\rangle=\cos\left(\frac{\Theta}{2}\right)|+\rangle+e^{i\Phi}\,\sin\left(\frac{\Theta}{2}\right)|-\rangle,
\end{equation}
where $0<\Theta<\pi$ and $0<\Phi<2\pi$ represent the qubit state on the surface of the Bloch sphere. The postselection state for the system is defined as
\begin{equation}
|f\rangle=\sin\left(\frac{\Theta}{2}+\theta\right)|+\rangle-e^{i(\Phi+2\phi)}\,\cos\left(\frac{\Theta}{2}+\theta\right)|-\rangle,
\end{equation}
where $\theta$ and $\phi$ are angles representing the deviation (in both angular directions on the Bloch sphere, see Fig.~\ref{Sphere}) from the state orthogonal to $|i\rangle$.
The probability of the postselection on $|f\ra$ and the weak value take the form,
\begin{eqnarray}
&\gamma \approx  |\langle f|i\rangle|^2 & =\cos^2\phi\,\sin^2\theta+\sin^2(\Theta+\theta)\,\sin^2\phi,\\
&\mbox{Re}\left(A_w\right)=&\frac{\sin\theta\,\sin(\Theta+\theta)}{|\langle f|i\rangle|^2},\\
&\mbox{Im}\left(A_w\right)=&-\frac{\sin\Theta\,\sin(\Theta+2\,\theta)\,\sin(2\phi)}{2\,|\langle f|i\rangle|^2}.
\end{eqnarray}

Note that the angles $\theta$ and $\phi$ are the respective generators of the real and the imaginary part of the weak value. For example, a pure imaginary weak value can be obtained making $\theta=0$, so $\gamma=\sin^2\Theta\,\sin^2\phi$ and $A_w=-i\,\cot\phi$. In order to maximize $\gamma$ we choose $\Theta=\pi/2$, so the results are $|i\rangle=\left[|+\rangle+e^{i\Phi}|-\rangle\right]/\sqrt{2}$ and $|f\rangle=\left[e^{-i\phi}|+\rangle-e^{i\Phi+i\phi}|-\rangle\right]\sqrt{2}$. Finally, in the small angle approximation, $\gamma\approx\phi^2$ and $A_w\approx-i/\phi$, making $|\langle f|i\rangle\,\mbox{Im}(A_w)|\approx 1$. On the other side, a pure real weak value can be obtained making $\phi=0$, so $\gamma=\sin^2\theta$, and $A_w=\sin(\Theta+\theta)/\sin\theta$. In order to make $A_w$ large for this case, we choose again $\Theta=\pi/2$. The preselection is identical to the former case $|i\rangle=\left[|+\rangle+e^{i\Phi}|-\rangle\right]/\sqrt{2}$, and the postselection takes the form $|f\rangle=\sin\theta\left[|+\rangle+e^{i\Phi}|-\rangle\right]/\sqrt{2}+\cos\theta\left[|+\rangle-e^{i\Phi}|-\rangle\right]/\sqrt{2}$. Taking the small angle approximation for this case we obtain a similar result to the pure imaginary weak value case, $\gamma\approx\theta^2$, $A_w\approx1/\theta$, and $|\langle f|i\rangle\,\mbox{Re}(A_w)|\approx 1$.

\begin{center}
\begin{figure}[t]
\includegraphics[scale=0.55]{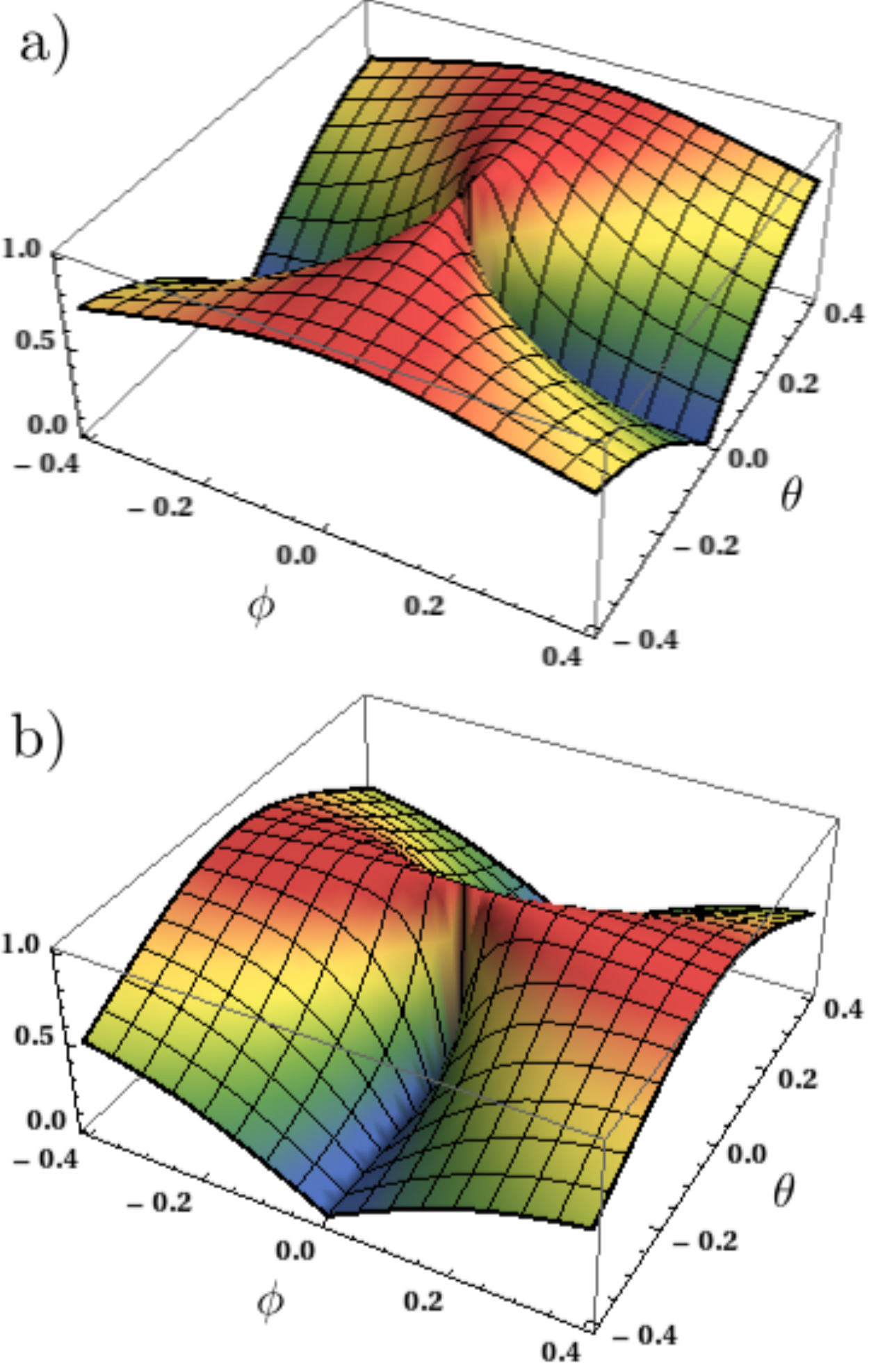}
\caption{\label{plots}$|\langle f|i\rangle\,\mbox{Re}(A_w)|$ (a) and $|\langle f|i\rangle\,\mbox{Im}(A_w)|$ (b) as a function of $\theta$ and $\phi$, for $\Theta=\pi/2$ (see text for details). Both functions are upper bounded by unity.}
\end{figure}
\end{center}
Setting $\Theta=\pi/2$ turns out to be the best choice for practical purposes. We  therefore calculate the products $|\langle f|i\rangle\,\mbox{Re}(A_w)|$ and $|\langle f|i\rangle\,\mbox{Im}(A_w)|$ for such a choice, and plot them in Fig.~\ref{plots}. It is shown in Fig.~\ref{plots}(a) that the largest value $|\langle f|i\rangle\,\mbox{Re}(A_w)|=1$ is closest reached if $\phi=0$ and $|\theta|\ll 1$, which corresponds to a pure real weak value and an almost orthogonal postselection. Similarly, Fig.~\ref{plots}(b) shows that $|\langle f|i\rangle\,\mbox{Im}(A_w)|\approx1$ only if $\theta=0$ and $|\phi|\ll 1$, defining a pure imaginary weak value with almost orthogonal postselection.



\begin{thebibliography}{10}
\bibitem{note}

In this case the interaction takes the form $U=\exp(-ik\hat{A}\hat{x})$, where tracking the beam's transverse position corresponds to an imaginary weak value. We also note that while Knee and Gauger treat imaginary weak values in their Appendix B, they do not consider the addition of technical noise in this case, and further compare the Fisher information of $k$ in the imaginary weak value case against a standard strategy measurement of $x$ (equivalent to not having a lens in Fig. 1(a)), which are different parameters.
\end{thebibliography}
\end{document}